\documentclass[reprint,amsmath,superscriptaddress,amssymb,aps,pra,twocolumn,showpacs,showkeys]{revtex4-1}
\usepackage{graphicx}
\usepackage{bm}
\usepackage{hyperref}
\usepackage{braket}
\usepackage[english]{babel}

\newcommand{\him}[1]{{\ensuremath{\rm #1}}}

\begin{document}
\preprint{APS/123-QED}
\title{Multi-photon subtracted thermal states: description, preparation and reconstruction}
\author{Yu. I. Bogdanov}
\affiliation{Institute of Physics and Technology, Russian Academy of Sciences, 117218, Moscow, Russia}
\affiliation{National Research Nuclear University "MEPHI", 115409, Moscow, Russia}
\affiliation{National Research University of Electronic Technology MIET, 124498, Moscow, Russia}
\author{K. G. Katamadze}\email{k.g.katamadze@gmail.com}
\affiliation{Institute of Physics and Technology, Russian Academy of Sciences, 117218, Moscow, Russia}
\affiliation{M. V. Lomonosov Moscow State University, 119991, Moscow, Russia}
\author{G. V. Avosopyants}
\affiliation{Institute of Physics and Technology, Russian Academy of Sciences, 117218, Moscow, Russia}
\affiliation{National Research University of Electronic Technology MIET, 124498, Moscow, Russia}
\affiliation{M. V. Lomonosov Moscow State University, 119991, Moscow, Russia}
\author{L.~V. ~Belinsky}
\affiliation{Institute of Physics and Technology, Russian Academy of Sciences, 117218, Moscow, Russia}
\affiliation{National Research University of Electronic Technology MIET, 124498, Moscow, Russia}
\author{N. A. Bogdanova}
\affiliation{Institute of Physics and Technology, Russian Academy of Sciences, 117218, Moscow, Russia}
\affiliation{National Research University of Electronic Technology MIET, 124498, Moscow, Russia}
\author{A. A. Kalinkin}
\affiliation{M. V. Lomonosov Moscow State University, 119991, Moscow, Russia}
\author{S. P. Kulik}
\affiliation{M. V. Lomonosov Moscow State University, 119991, Moscow, Russia}

\begin{abstract}
	We present a study of optical quantum states generated by subtraction of photons from the thermal state. Some aspects of their photon number and quadrature distributions are discussed and checked experimentally. We demonstrate an original method of up to ten photon subtracted state preparation with use of just one single-photon detector. All the states where measured with use of balanced homodyne technique, and the corresponding density matrices where reconstructed. The fidelity between desired and reconstructed states exceeds $ 99\%
$.
Combined with homodyne detection it can also be used for precise measurement of high-order autocorrelation functions.
\end{abstract}
\maketitle
\section{Introduction}
	Preparation and measurement of various quantum states of light are the keystones of quantum optics. So far only a few classes of quantum states were available for experimental research. Among them there are displaced and squeezed states, the first few Fock states, Schr\"{o}dinger cat states etc. One of them, namely the thermal state, plays a special role. On the one hand, it is an easy-to-prepare state, but on the other, it supports classical correlations and can be used as a test site area for effects based on classical or quantum correlations.
	
	It is worthy to mention that the first pioneer experiment in quantum optics is considered to be the work by Hanbury Brown and Twiss \cite{BROWN1956}, who investigated correlations in thermal light by means of a beam splitter and a pair of detectors, outputs of which are analyzed with a coincidence circuit. Since then thermal states have been used in many applications including ghost imaging \cite{Gatti2004,Ferri2005,Valencia2005}, quantum illumination \cite{Lloyd2008}, and \textquotedblleft thermal laser\textquotedblright \cite{Chekhova1996b}. Schmidt-like correlations \cite{Bobrov2013} and HOM-interference \cite{Liu2013} were also observed for thermal states.
In the present paper we study a family of thermal states modified by multiphoton subtraction.
	
	Photon addition and subtraction is of great interest in quantum optics, because it provides a tool for direct tests of basic commutation relations \cite{Parigi2007}, enables Schr\"{o}dinger cat \cite{Wenger2004} and other non-Gaussian quantum state preparation. It can also be used for probabilistic linear no-noise amplification \cite{Xiang2010}. One- and two-photon subtracted thermal states were demonstrated for the first time in \cite{Zavatta2008}. Next up to eight-photon subtracted thermal state was prepared with use of photon number resolved detectors \cite{Allevi2010, Zhai2013}.
	
	In the present work we analyze the quadrature distribution of multiphoton subtracted thermal states ({MPSTS}) both theoretically and experimentally. The text is organized as follows. In the section \ref{sec:PS} we introduce a universal approach for photon number distribution calculation of arbitrary multiphoton subtracted quantum states, which is based on generating functions. Using this technique, we find photon number and quadrature distributions for MPSTS. In the section \ref{sec:Prep} we describe an experimental technique of MPSTS preparation with using just one non-photon-number-resolving single-photon detector. In the section \ref{sec:Rec} we show, how one can apply the model of MPSTS, found in previous section, to the density matrix reconstruction from the quadrature measurements. Finally, the experimental results are presented and discussed in the section \ref{sec:Results}. The utilization of photon subtraction of the thermal state for precise interferometric phase measurements was recently reported \cite{Rafsanjani2017}.
	
\section{Photon subtracted states}\label{sec:PS}
	Photon number distribution $ P(n) $ is a key characteristic of any quantum state of light. Any particular distribution corresponds to its generating function  $ G(z) $, which can be defined by equation:
	\begin{equation}\label{eq:Pn}
		G(z)=\sum_{n}P(n) z^{n},\quad P(n)=\frac{G^{(n)}(0)}{n!},
	\end{equation}
	where $ G^{(n)} $ is an $ n $-th order derivative. Properties of the annihilation operator and renormalization conditions lead us to the simple description of photon subtraction \cite{Bogdanov2016b}:
	\begin{equation}\label{eq:subtr1}
		G_{1}(z)=\frac{G^{(1)}(z)}{\mu},
	\end{equation}
	where $ G_{1}(z) $  is the generating function, which corresponds to the photon subtracted state and $ \mu $ is a mean photon number of the initial state. Applying (\ref{eq:subtr1}) $ k $  times, one can find the generating function for the $ k $-photon subtracted state:
	\begin{equation}\label{eq:subtrk}
		G_{k}(z)=\frac{G^{(k)}(z)}{\mu\mu_{1}\cdots\mu_{k-1}},
	\end{equation}
	where $ \mu_{k} $ is a mean photon number of $ k $-photon subtracted state.
	
	Equations (\ref{eq:subtr1}) and (\ref{eq:subtrk}) can be used for calculation of the distribution $ P(n) $ (\ref{eq:Pn}) as well as for the  $ m $-th order correlation function calculation:
	\begin{equation}\label{eq:gm}
		g^{(m)}=\frac{G^{(m)}(1)}{\mu^{m}}=\frac{\mu_{1}\mu_{2}\cdots\mu_{m-1}}{\mu^{m-1}}, m=2, 3,\ldots
	\end{equation}
	Let's consider several examples.
	\subsection{Fock state}
		The photon number distribution of the Fock state $ \ket{m} $ is $ P(n)=\delta_{m,n} $ and its generating function $ G(z)=z^{m} $. After photon subtraction (\ref{eq:subtr1}) it transforms to $ G_{1}(z)=z^{m-1} $, which corresponds to the state $ \ket{m-1} $.
	\subsection{Coherent state}
		A coherent state can be written in the Fock basis as $ \ket{\alpha}={e^{-\frac{|\alpha|^{2}}{2}}\sum\limits_{n=0}^{\infty}\frac{\alpha^{n}}{\sqrt{n!}}\ket{n}} $, so it’s photon number has a Poisson distribution $ P(n)=e^{-|\alpha|^{2}}\frac{|\alpha|^{2n}}{{n!}} $ with the mean photon number $ \mu=|\alpha|^{2} $ so the generating function turns $ G(z)=e^{\mu(z-1)} $. Applying photon subtraction (\ref{eq:subtr1}) one can verify that $ G_{1}(z)=G(z) $, which means that coherent state doesn't change under photon subtraction.
	\subsection{Squeezed vacuum}
		The photon number distribution of the  squeezed vacuum state $ \hat{S}(\xi)\ket{0} $ is \cite{Scully2001}
		\begin{equation}\label{eq:Psq}
			\begin{split}
				&P(2n)=\frac{1}{\cosh(|\xi|)}\frac{(2n)!}{(n!)^{2}}\left(\frac{1}{2}\tanh(|\xi|)\right)^{2n},\\
				&P(2n+1)=0, n=0,1,\ldots
			\end{split}
		\end{equation}
		Its generating function equals
		\begin{equation}
			G(z)=\frac{1}{\cosh(|\xi|)\sqrt{1-z^{2}\tanh^{2}(|\xi|)}},
		\end{equation}
		and its mean photon number is $ \mu=G^{(1)}(1)=\sinh^{2}(|\xi|) $.
		
		Using this approach, one can, for example, calculate a high-order correlation function of squeezed vacuum:
		\begin{equation}\label{geq:sq}
			g^{(n)}=\frac{n!}{2^{n}}\sum\limits_{k=0}^{\lfloor{n/2}\rfloor}\frac{(2n-2k)!}{k!(n-k)!(n-2k)!}\left(\frac{1}{\sinh^{2}(|\xi|)}\right)^{k},
		\end{equation}
		where $ \lfloor{\ldots}\rfloor $ is the floor function.
	\subsection{Thermal state}
	
			\begin{figure*}[t]
			\center{\includegraphics[width=0.9\textwidth]{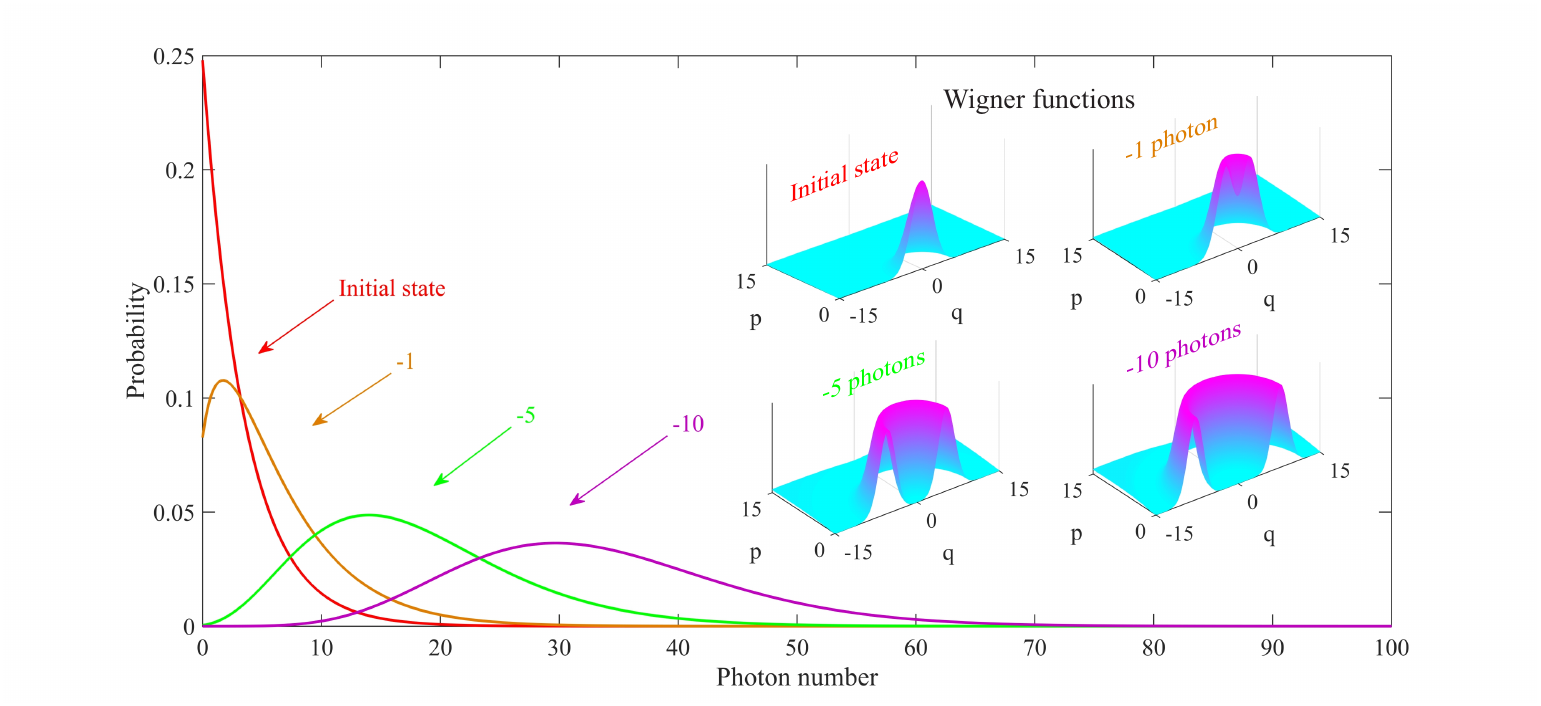}}
			\caption{Photon number distributions and Wigner functions for initial thermal state and $ k $-photon subtracted thermal states with $ k=1,5,10 $.}
			\label{fig:wign}
		\end{figure*}

		The density matrix of a thermal state has a well-known diagonal form:
		\begin{equation}\label{eq:rho}
			\hat{\rho}=\sum\limits_{n=0}^{\infty}P(n)\ket{n}\bra{n},
		\end{equation}
		where $ P(n)={\mu^{n}}/{(1+\mu)^{n+1}} $ is a Bose-Einstein distribution. This distribution is a particular case of compound Poisson distribution
		\begin{equation}\label{eq:Pps}
			P_{\mu,a}(n)=\frac{\Gamma(a+n)}{\Gamma(a)}\frac{\mu^{n}}{a^{n}n!}\frac{1}{\left(1+\mu/a\right)^{n+a}}.
		\end{equation}
		This distribution have two parameters: the mean photon number $ \mu $ and coherence parameter $ a $. At $ a=1 $ equation (\ref{eq:Pps}) turns into the Bose-Einstein distribution, and at $ a\longrightarrow\infty $ (\ref{eq:Pps}) turns into the ordinary Poisson distribution. This distribution describes a multimode thermal state, where $ a $ is the number of modes \cite{Mandel1995}.
		
		It can be shown, that the same distribution applies also to the single-mode multiphoton-subtracted thermal state \cite{Allevi2010, Zhai2013, Bogdanov2016b}.
		
		Its generating function equals:
		\begin{equation}\label{eq:Gps}
		G(z)=\left[1+\frac{(1-z)\mu}{a}\right]^{-a}.
		\end{equation}
		Using (\ref{eq:subtr1}) one can show that photon subtraction conserves the type of the distribution (\ref{eq:Pps}), but changes the values of parameters $ a $ and $ \mu $ as follows: $ a_{1}=a+1 $, $ \mu_{1}=\mu\frac{a+1}{a} $. Using these iterative relations we can see that a thermal state with the initial parameters $ \mu_{0} $ and $ a_{0}=1 $  after subtraction of $ k $ photons transforms into the state (\ref{eq:rho}), (\ref{eq:Pps}) with parameters
		\begin{equation}\label{eq:amu}
		a_{k}=k+1,\quad \mu_{k}=\mu_{0}(k+1).
		\end{equation}
		It is rather counterintuitive that the mean photon number increases after photon subtraction procedure. This can be explained as follows. Probabilistic photon subtraction can be realized by means of a low-reflective beam splitter combined with a single-photon detector in the reflection channel, which clicks if the photon annihilation takes place \cite{Wenger2004}. As the reflection of the beam splitter is very weak, most of the time there are no detector clicks. However, when a photon is detected it results in the following: 1. there is one less photon after the beam splitter than before; 2. the number of photons before the beam splitter was greater (on the average) than the mean. In our case the second factor is much greater than the first one. Let us mention, that for coherent states with Poisson photon distributions these two factors compensate each other, so the photon subtraction doesn't change the mean photon number.
		
		This peculiar behavior can be effectively used as probabilistic amplification due to photon subtraction, which enables higher phase sensitivity in thermal field interferometry \cite{Rafsanjani2017}. 		
		In contrast, ordinary losses only decrease $ \mu $ and conserve $ a $.
			
		Using (\ref{eq:gm}), we can show that the correlation function of a $ k $-photon subtracted thermal state equals
		\begin{equation}\label{eq:g2}
			g^{2}=1+\frac{1}{a}=1+\frac{1}{k+1}.
		\end{equation}
		This equation is similar to the correlation function for multi-mode thermal state \cite{Mandel1995}.

		Photon number distributions for several photon-subtracted thermal states as well as their Wigner functions are shown in Fig. \ref{fig:wign}. Following the procedure of photon subtraction, the initial Gaussian function transforms to a ring-shaped non-Gaussian function, whose radius is approximately proportional to  $ \sqrt{\mu_{k}} $. The non-Gaussianity of MPSTS has been studied recently \cite{Ghiu2014}.

We can also find a quadrature distribution of MPSTS:
		\begin{equation}\label{eq:quadrature}
			P_{\mu,a}(q)=\sum_{n=0}^{\infty}{P_{\mu,a}(n)\left|\varphi_n(q)\right|^2},
		\end{equation}	
where $\varphi_n(q)$ are the Hermit eigenfunctions of harmonic oscillator:
\begin{equation}\label{eq:ocsillator}
	\varphi_n(q)=\frac{H_k(q)}{(2^k k! \sqrt{\pi})^{1/2}} e^{-x^2/2},
\end{equation}
$H_k$ are Hermite polynomials.

\begin{figure*}
			\center{

  \includegraphics[width=0.74\textwidth]{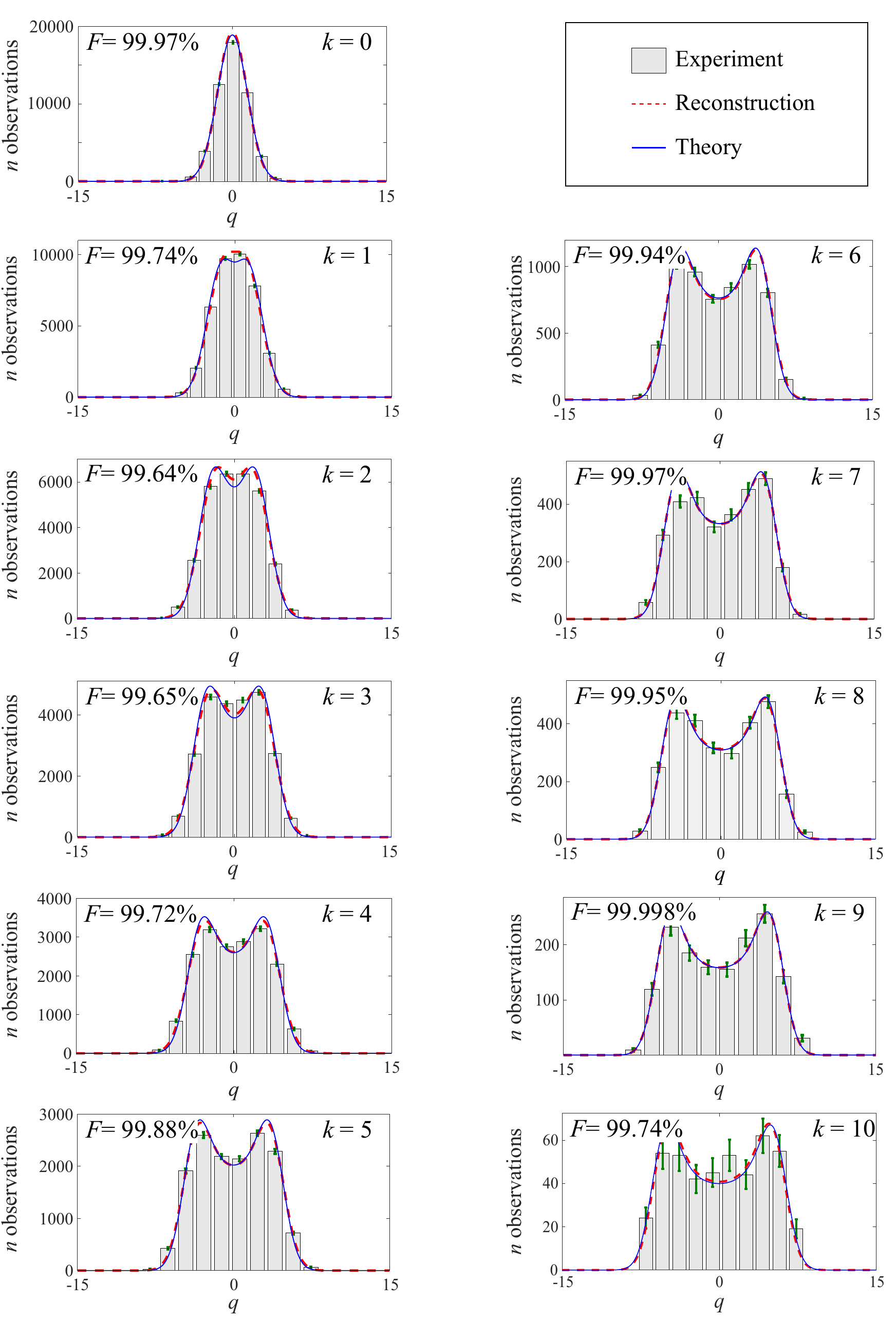}

  }
			\caption{Quadrature distributions $ P(q) $ for the $k$-photon subtracted thermal states with $k=0-10$. Experimental data are plotted as histograms with statistical errors, the MLE fit is plotted as a red dashed line and theoretical distribution as a blue solid line.}
			\label{fig:quadr}
		\end{figure*}

		The quadrature distributions $ P(q) $ for 0--10-photon subtracted thermal states are shown in Fig.~ \ref{fig:quadr}. It can be calculated that the variance  $ \sigma^{2} $ and the 
kurtosis  $ K\equiv\overline{(q-\bar{q})^{4}}/\sigma^{4}
$ relates to photon distribution parameters $ a $ and $ \mu $ as
		\begin{equation}\label{eq:amuexp}
			\sigma^{2}=\mu+\frac{1}{2},\quad K=3-6\left(\frac{\mu}{2\mu+1}\right)^{2}\frac{a-1}{a}.
		\end{equation}		
		These relations can be used for quick estimation of $ a $ and $ \mu $ from homodyne measurements.	
	
\section{Experiment}\label{sec:Prep}
\begin{figure}[t]
		\center{\includegraphics[width=0.5\textwidth]{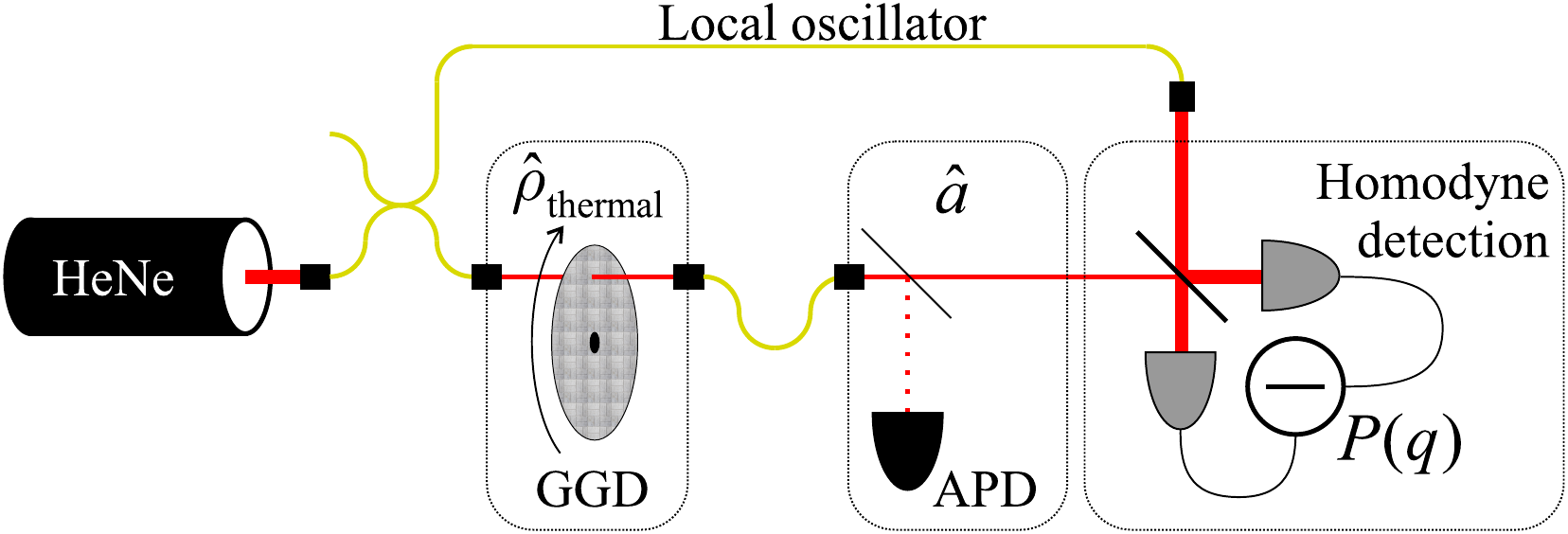}}
		\caption{Experimental setup. Thermal state $ \rho_{\text{thermal}} $ is prepared from a HeNe laser radiation by randomizing its phase and amplitude in a rotating ground glass disk (GGD) \cite{Martienssen1964,Arecchi1965}. Photon subtraction $ \hat{a} $ is realized with a low-refractive beam splitter combined with a single-photon APD detector. The quadrature distribution of prepared state is measured with the homodyne detection technique \cite{Leonhardt1995}.}
		\label{fig:setup}
	\end{figure}

	The sketch of the experimental setup is shown in Fig.~\ref{fig:setup}. The HeNe cw laser radiation at the wavelength of $ 633$~nm is coupled with a single-mode fiber and asymmetrically split into two channels. The main part of radiation serves as a local oscillator and the leftover part is utilized for quantum state preparation. The initial quasi-thermal state is prepared by passing the laser beam through the rotating ground glass  disk \cite{Martienssen1964, Arecchi1965}. The corresponding coherence time of $ \tau_{coh}=40$~\him{\mu s}  approximately equals the time it takes for a grain of the disk to cross the laser beam and can be tuned by the disk displacement and its speed variation. For the single spatial mode selection, the scattered radiation is passed again through  the single-mode fiber.   Conditional photon subtraction is realized by a beam splitter with reflectivity $ r=1\%
$
combined with an APD single photon detector Laser Components COUNT-100C-FC with 100~Hz dark counts and a 50~ns dead time, placed in the reflection channel \cite{Wenger2004}. Finally, the quadrature distribution of the obtained photon subtracted thermal state is measured with the homodyne technique \cite{Leonhardt1995}. We used a commercial balanced homodyne detector Thorlabs PDB450A with a 100 kHz bandwidth and a 78\%
quantum efficiency. The Wigner functions of measured states are axially symmetrical (see Fig.~\ref{fig:wign}), so the homodyne phase isn't varied.

	\begin{figure}
		\center{\includegraphics[width=0.45\textwidth]{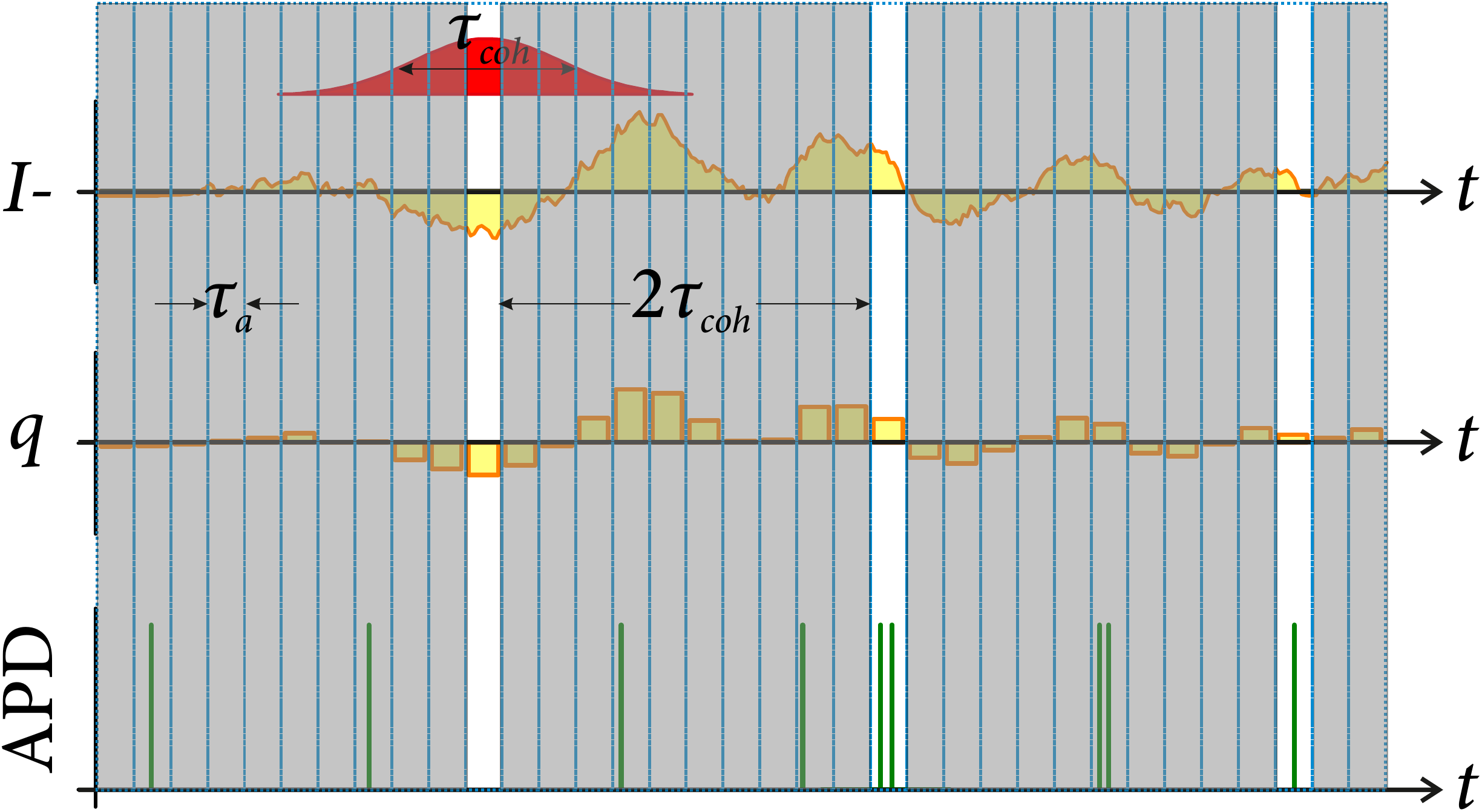}}
		\caption{Experimental data processing (qualitative picture). The quadrature values $q$ (center plot) obtained by the difference photo current $I_-$ (top plot)  integration over the acquisition interval $\tau_a$. This interval is smaller than the width $\tau_{coh}$ of the natural time mode $\psi(t)$ (red bell-shaped plot), which can be defined by the correlation function measurement. So the measured time mode is rectangular shaped with the width $\tau_a$. Every APD photo count (bottom plot) corresponds to the photon subtraction. Time bins periodically separated by the  $2\tau_{coh}$ where selected for the further quantum state reconstruction}
		\label{fig:processing}
	\end{figure}

	The main difference of our setup from the others \cite{Zavatta2008, Allevi2010, Zhai2013} is a cw regime, which allow us to use just one APD detector for a multiple photon subtraction. It can be done as follows (see Fig.~\ref{fig:processing}).
%
%
The natural bell-shaped time mode $\psi(t)$ of the pseudo-thermal light can be characterized by the correlation function $g^{(2)}(t)$ with the width $\tau_{coh}=40$~\him{\mu s}. The measured quadrature value $q$ is obtained by the difference photo current $I_-$ integration over the averaging time $ \tau_a$: $q\propto\int_{\tau_a}I_-(t)\psi(t)dt$ \cite{Kumar2012}.   Choosing the acquisition time $ \tau_a=12\ \mu\mbox{s} < \tau_{coh}$ we cut the central part of the mode $\psi(t)$. So our measured mode is now rectangle-shaped with the width $ \tau_a$. Every photo count registered inside this $ \tau_a$-interval corresponds to the photon subtraction from this measured mode. If the APD dead time $\tau_{d}=50\ \mbox{ns}\ll \tau_a$, we can register several photo counts inside the acquisition interval, which corresponds to multiple photon subtraction. To avoid any interbin correlations we select the bins periodically separated by $2\tau_{coh}$. We should note, that it is possible to use the data from all the bins, it significantly increases the sample size, but the measured values become statistically dependent so the $\chi^{(2)}$-test (see next section) can no longer be applied.

The multiple photon subtraction method is a quite similar (up to space-time exchange) to the principle of operation of photon number resolved detector, based on the APD array \cite{Yamamoto2007}, where several photons in one spatial mode can be independently detected by different APD's, placed in the different points of the initial spatial mode area.
Two-photon subtracted thermal states were recently realised using this technique \cite{Rafsanjani2017}.
It can also be used in other cw experiments, for example for modification of the squeezed vacuum states \cite{Neergaard-Nielsen2006}. The necessary condition $\tau_{coh}\gg\tau_{d}\approx50\ \mbox{ns}$ can be satisfied for example, in case of 2~MHz narrow-band spontaneous parametric down-conversion \cite{Rielander2014}.

The measured conditional quadrature distributions were used to reconstruct the prepared quantum states of light.
\section{Reconstruction}\label{sec:Rec}
	\begin{figure}[t]
		\center{\includegraphics[width=0.45\textwidth]{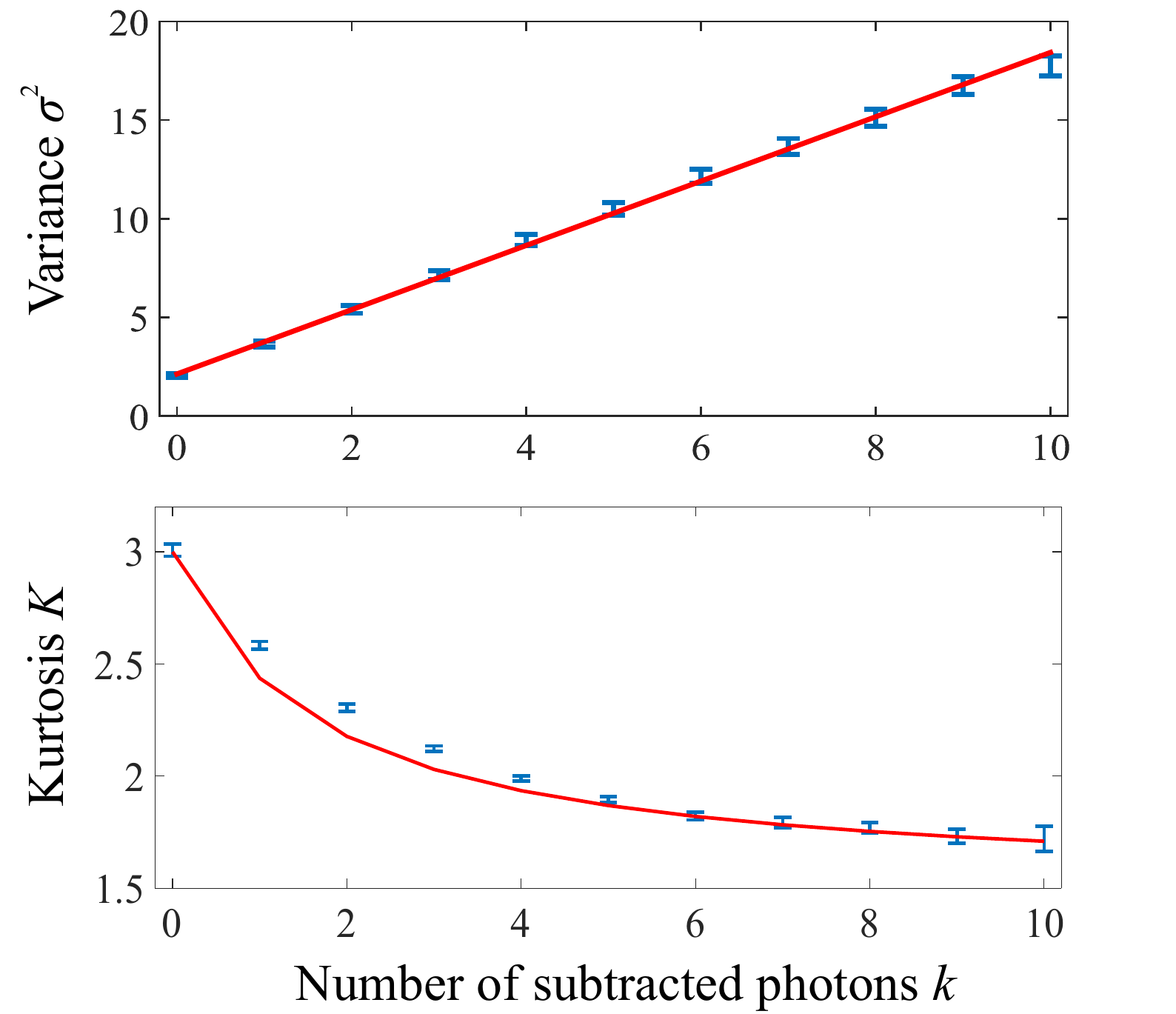}}
		\caption{Dependency of the quadrature distribution variance $ \sigma $ and kurtosis $ K $ on the number of subtracted photons. Dots corresponds to experimental values and lines -- to theoretical predictions (\ref{eq:amu}, \ref{eq:amuexp}).}
		\label{fig:SigmaK}
	\end{figure}

	An easy way to estimate the quantum state  (\ref{eq:rho}, \ref{eq:Pps}) from experimental quadrature data is based on the relations (\ref{eq:amuexp}). The quadrature variance $\sigma^2$ and kurtosis $K$ vs. the number of subtracted photons are plotted in Fig.~\ref{fig:SigmaK} and the experimental dots lie close to the theory curves (\ref{eq:amu}, \ref{eq:amuexp}).  However, for more accurate reconstruction we used the maximum likelihood estimation (MLE). Typically, the MLE is used to reconstruct the density matrix of the state $\hat{\rho}=\sum\limits_{n,m=0}^{N}\rho_{n,m}\ket{n}\bra{m} $, where $ N $ is a limit of maximum photon number \cite{Lvovsky2004}.

	This model is quite general, but not optimal, because the number of estimated parameters is too large; the corresponding problem is ill-conditioned and requires a lot of computing power. Therefore, it gives rather low precision of estimates. For a considerable set of experimentally available quantum states of light, the model based on the basis of displaced squeezed Fock states and root approach can be used for significant decrease the number of estimated parameters \cite{Bogdanov2016}.
	
	However, a simpler model based on the compound Poisson photon number distribution (\ref{eq:rho}, \ref{eq:Pps}) is sufficient for the purposes of this paper. We just need to fit measured quadrature distribution $P(q)$ with the model distribution $P_{\mu,a} (q)$ (\ref{eq:quadrature}) and find the values of $a$ and $\mu$, which maximize the likelihood function. To account the homodyne detection efficiency $\eta$ we smooth the model distribution $P_{\mu,a} (q)$ with a Gaussian function $e^{-q^2 \eta/(1-\eta)}$ \cite{Leonhardt1995}. Our model exploits only two real parameters, so high precision quantum state estimation can be performed. However every time one should check whether the estimated function
$P_{\mu,a} (q)$ is a good fit for the experimental data $P(q)$. This verification was done with the usual  $ \chi^{2}$--test. The significance level was higher than 0.01 for all of the prepared and measured states. In Fig.~\ref{fig:quadr} one can see that the dashed red lines, obtained by MLE, are indeed a good fit for the experimental quadrature data, plotted as a histograms, and lie close to the solid blue lines, which correspond to the state (\ref{eq:rho},~\ref{eq:Pps}) with theoretically predicted values of $a$ and $\mu$.

\section{Results}\label{sec:Results}
	
\begin{figure}[t]
		\center{\includegraphics[width=0.40\textwidth]{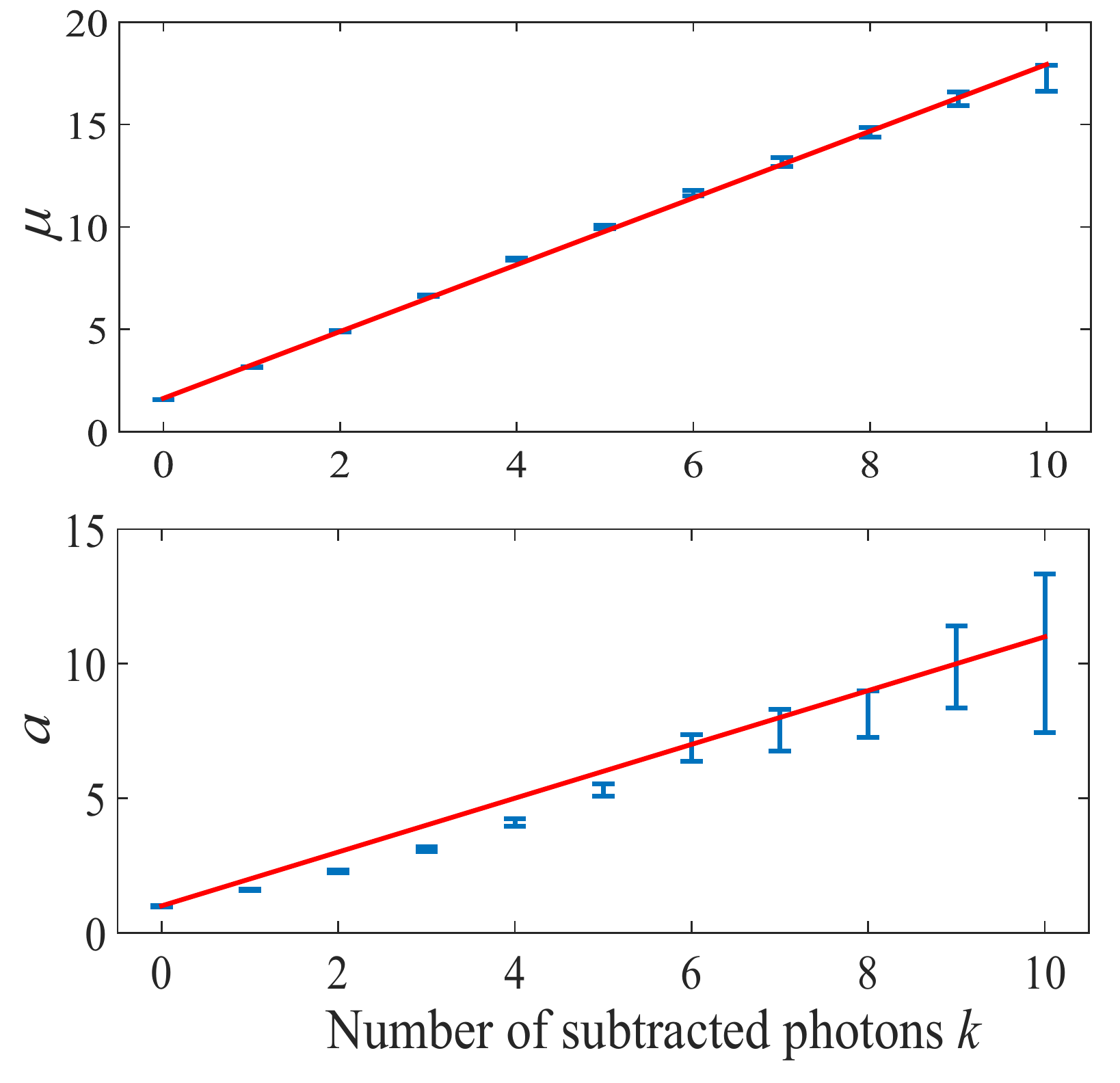}}
		\caption{Dependency of the mean photon number $ \mu $ and coherence parameter $ a $ on the number of subtracted photons. Dots corresponds to experimental values and lines -- to theoretical predictions (\ref{eq:amu}).}
		\label{fig:error}
	\end{figure}

	Eleven different quantum states were prepared, measured and reconstructed, namely the initial thermal state with the mean photon number  $ \mu=1.63 $, and $ k $-photon subtracted thermal states, where $ k=1,\ldots,10 $. The estimated values of $ a $ and $ \mu $ are plotted in Fig.~\ref{fig:error}. Lines correspond to the predicted values of the parameters (\ref{eq:amu}). As follows from the figure, experimental results are in the good agreement with theoretical predictions.
		Error bars of the estimated parameters were calculated using the Fisher information matrix. Large uncertainties for $ k=9,10 $ are due to the small volume of the sampled data (just 1500 and 450 points).
	
	We should note that in spite of the theory of photon subtraction predicts integer values of the parameter $a$ (\ref{eq:amu}) our model allows for real values of $a$ (\ref{eq:Pps}), which enables better fit of the experimental data. Such quantum states can be interpreted as a mixture of states with different numbers of subtracted photons. For example, one photo count may caused both by the photon subtraction and by the dark (or background) noise. So, the selection of events, corresponding to one photo count gives a mixture of the initial and one-photon-subtracted states.
	
	It's worth noting that all the experimental non-idealities such as APD dark counts, limited quantum efficiency and so on, do not cause significant deviations from the simple theory predictions.	We estimate the agreement between theoretical and experimental density matrices by calculating the fidelity:
	\begin{equation}\label{fid}
		F(\hat{\rho}_{th},\hat{\rho}_{exp})=\left(Tr\left(\sqrt{\sqrt{\hat{\rho}_{th}}\hat{\rho}_{exp}\sqrt{\hat{\rho}_{th}}}\right)\right)^{2}
	\end{equation}
	For all the measured states the fidelity is higher than $ 99\%
$. The calculated values of fidelity are also indicated in Fig.~\ref{fig:quadr}.
%

	We should also mention that the obtained fidelity values are rather high in spite of the estimated values of parameter $a$ deviating from values predicted by the theory (Fig.~\ref{fig:error}). This means that the $a$ value is more sensitive to the changes in quantum state than the fidelity.
\section{Conclusion}
	Quadrature distributions of photon-subtracted thermal states have been studied both theoretically (based on generating function approach (\ref{eq:subtr1})) and experimentally. Simple equations (\ref{eq:amuexp}) for quadrature distributions of MPSTS have been found. Up to ten-photon subtracted states have been experimentally realized with a single APD by means of a long coherence time of the initial thermal state (Fig.~\ref{fig:setup},~\ref{fig:processing}).
Applying MLE and using fitting functions with two real parameters (\ref{eq:quadrature}) we were able to reconstruct selected quantum states with high accuracy  by measuring quadrature distributions
%
%
This simple model fits rather  well the experimental data shown in Fig.~\ref{fig:quadr}. The estimated states are in a good agreement (fidelity $>99\%
$) with the theoretical prediction.

\begin{acknowledgments}
	This work was supported by Russian Foundation of Basic Research (project no: 14-02-00749 A), by the Grant of President of Russian Federation no: MK-5860.2016.2., and by the Program of the Russian Academy of Sciences in fundamental research.
	
\end{acknowledgments}


\begin{thebibliography}{28}%
\makeatletter
\providecommand \@ifxundefined [1]{%
 \@ifx{#1\undefined}
}%
\providecommand \@ifnum [1]{%
 \ifnum #1\expandafter \@firstoftwo
 \else \expandafter \@secondoftwo
 \fi
}%
\providecommand \@ifx [1]{%
 \ifx #1\expandafter \@firstoftwo
 \else \expandafter \@secondoftwo
 \fi
}%
\providecommand \natexlab [1]{#1}%
\providecommand \enquote  [1]{``#1''}%
\providecommand \bibnamefont  [1]{#1}%
\providecommand \bibfnamefont [1]{#1}%
\providecommand \citenamefont [1]{#1}%
\providecommand \href@noop [0]{\@secondoftwo}%
\providecommand \href [0]{\begingroup \@sanitize@url \@href}%
\providecommand \@href[1]{\@@startlink{#1}\@@href}%
\providecommand \@@href[1]{\endgroup#1\@@endlink}%
\providecommand \@sanitize@url [0]{\catcode `\\12\catcode `\$12\catcode
  `\&12\catcode `\#12\catcode `\^12\catcode `\_12\catcode `\%12\relax}%
\providecommand \@@startlink[1]{}%
\providecommand \@@endlink[0]{}%
\providecommand \url  [0]{\begingroup\@sanitize@url \@url }%
\providecommand \@url [1]{\endgroup\@href {#1}{\urlprefix }}%
\providecommand \urlprefix  [0]{URL }%
\providecommand \Eprint [0]{\href }%
\providecommand \doibase [0]{http://dx.doi.org/}%
\providecommand \selectlanguage [0]{\@gobble}%
\providecommand \bibinfo  [0]{\@secondoftwo}%
\providecommand \bibfield  [0]{\@secondoftwo}%
\providecommand \translation [1]{[#1]}%
\providecommand \BibitemOpen [0]{}%
\providecommand \bibitemStop [0]{}%
\providecommand \bibitemNoStop [0]{.\EOS\space}%
\providecommand \EOS [0]{\spacefactor3000\relax}%
\providecommand \BibitemShut  [1]{\csname bibitem#1\endcsname}%
\let\auto@bib@innerbib\@empty
\bibitem [{\citenamefont {Brown}\ and\ \citenamefont
  {Twiss}(1956)}]{BROWN1956}%
  \BibitemOpen
  \bibfield  {author} {\bibinfo {author} {\bibfnamefont {R.~H.}\ \bibnamefont
  {Brown}}\ and\ \bibinfo {author} {\bibfnamefont {R.~Q.}\ \bibnamefont
  {Twiss}},\ }\href@noop {} {\bibfield  {journal} {\bibinfo  {journal}
  {Nature}\ }\textbf {\bibinfo {volume} {177}},\ \bibinfo {pages} {27}
  (\bibinfo {year} {1956})}\BibitemShut {NoStop}%
\bibitem [{\citenamefont {Gatti}\ \emph {et~al.}(2004)\citenamefont {Gatti},
  \citenamefont {Brambilla}, \citenamefont {Bache},\ and\ \citenamefont
  {Lugiato}}]{Gatti2004}%
  \BibitemOpen
  \bibfield  {author} {\bibinfo {author} {\bibfnamefont {A.}~\bibnamefont
  {Gatti}}, \bibinfo {author} {\bibfnamefont {E.}~\bibnamefont {Brambilla}},
  \bibinfo {author} {\bibfnamefont {M.}~\bibnamefont {Bache}}, \ and\ \bibinfo
  {author} {\bibfnamefont {L.~A.}\ \bibnamefont {Lugiato}},\ }\href {\doibase
  10.1103/PhysRevLett.93.093602} {\bibfield  {journal} {\bibinfo  {journal}
  {Physical Review Letters}\ }\textbf {\bibinfo {volume} {93}},\ \bibinfo
  {pages} {093602} (\bibinfo {year} {2004})}\BibitemShut {NoStop}%
\bibitem [{\citenamefont {Ferri}\ \emph {et~al.}(2005)\citenamefont {Ferri},
  \citenamefont {Magatti}, \citenamefont {Gatti}, \citenamefont {Bache},
  \citenamefont {Brambilla},\ and\ \citenamefont {Lugiato}}]{Ferri2005}%
  \BibitemOpen
  \bibfield  {author} {\bibinfo {author} {\bibfnamefont {F.}~\bibnamefont
  {Ferri}}, \bibinfo {author} {\bibfnamefont {D.}~\bibnamefont {Magatti}},
  \bibinfo {author} {\bibfnamefont {A.}~\bibnamefont {Gatti}}, \bibinfo
  {author} {\bibfnamefont {M.}~\bibnamefont {Bache}}, \bibinfo {author}
  {\bibfnamefont {E.}~\bibnamefont {Brambilla}}, \ and\ \bibinfo {author}
  {\bibfnamefont {L.~A.}\ \bibnamefont {Lugiato}},\ }\href {\doibase
  10.1103/PhysRevLett.94.183602} {\bibfield  {journal} {\bibinfo  {journal}
  {Physical Review Letters}\ }\textbf {\bibinfo {volume} {94}},\ \bibinfo
  {pages} {183602} (\bibinfo {year} {2005})},\ \Eprint
  {http://arxiv.org/abs/0408021} {arXiv:0408021 [quant-ph]} \BibitemShut
  {NoStop}%
\bibitem [{\citenamefont {Valencia}\ \emph {et~al.}(2005)\citenamefont
  {Valencia}, \citenamefont {Scarcelli}, \citenamefont {D'Angelo},\ and\
  \citenamefont {Shih}}]{Valencia2005}%
  \BibitemOpen
  \bibfield  {author} {\bibinfo {author} {\bibfnamefont {A.}~\bibnamefont
  {Valencia}}, \bibinfo {author} {\bibfnamefont {G.}~\bibnamefont {Scarcelli}},
  \bibinfo {author} {\bibfnamefont {M.}~\bibnamefont {D'Angelo}}, \ and\
  \bibinfo {author} {\bibfnamefont {Y.}~\bibnamefont {Shih}},\ }\href {\doibase
  10.1103/PhysRevLett.94.063601} {\bibfield  {journal} {\bibinfo  {journal}
  {Physical Review Letters}\ }\textbf {\bibinfo {volume} {94}},\ \bibinfo
  {pages} {1} (\bibinfo {year} {2005})},\ \Eprint
  {http://arxiv.org/abs/0408001} {arXiv:0408001 [quant-ph]} \BibitemShut
  {NoStop}%
\bibitem [{\citenamefont {Lloyd}(2008)}]{Lloyd2008}%
  \BibitemOpen
  \bibfield  {author} {\bibinfo {author} {\bibfnamefont {S.}~\bibnamefont
  {Lloyd}},\ }\href {\doibase 10.1126/science.1160627} {\bibfield  {journal}
  {\bibinfo  {journal} {Science (New York, N.Y.)}\ }\textbf {\bibinfo {volume}
  {321}},\ \bibinfo {pages} {1463} (\bibinfo {year} {2008})}\BibitemShut
  {NoStop}%
\bibitem [{\citenamefont {Chekhova}\ \emph {et~al.}(1996)\citenamefont
  {Chekhova}, \citenamefont {Kulik}, \citenamefont {an~Penin}, \citenamefont
  {Prudkovskii}, \citenamefont {Penin},\ and\ \citenamefont
  {Prudkovskii}}]{Chekhova1996b}%
  \BibitemOpen
  \bibfield  {author} {\bibinfo {author} {\bibfnamefont {M.~V.~M.}\
  \bibnamefont {Chekhova}}, \bibinfo {author} {\bibfnamefont {S.~P.}\
  \bibnamefont {Kulik}}, \bibinfo {author} {\bibnamefont {an~Penin}}, \bibinfo
  {author} {\bibfnamefont {P.~A.}\ \bibnamefont {Prudkovskii}}, \bibinfo
  {author} {\bibfnamefont {A.~N.}\ \bibnamefont {Penin}}, \ and\ \bibinfo
  {author} {\bibfnamefont {P.~A.}\ \bibnamefont {Prudkovskii}},\ }\href
  {\doibase 10.1103/PhysRevA.54.R4645} {\bibfield  {journal} {\bibinfo
  {journal} {Physical review. A}\ }\textbf {\bibinfo {volume} {54}},\ \bibinfo
  {pages} {R4645} (\bibinfo {year} {1996})}\BibitemShut {NoStop}%
\bibitem [{\citenamefont {Bobrov}\ \emph {et~al.}(2013)\citenamefont {Bobrov},
  \citenamefont {Straupe}, \citenamefont {Kovlakov},\ and\ \citenamefont
  {Kulik}}]{Bobrov2013}%
  \BibitemOpen
  \bibfield  {author} {\bibinfo {author} {\bibfnamefont {I.~B.}\ \bibnamefont
  {Bobrov}}, \bibinfo {author} {\bibfnamefont {S.~S.}\ \bibnamefont {Straupe}},
  \bibinfo {author} {\bibfnamefont {E.~V.}\ \bibnamefont {Kovlakov}}, \ and\
  \bibinfo {author} {\bibfnamefont {S.~P.}\ \bibnamefont {Kulik}},\ }\href
  {\doibase 10.1088/1367-2630/15/7/073016} {\bibfield  {journal} {\bibinfo
  {journal} {New Journal of Physics}\ }\textbf {\bibinfo {volume} {15}},\
  \bibinfo {pages} {073016} (\bibinfo {year} {2013})},\ \Eprint
  {http://arxiv.org/abs/1210.3212} {arXiv:1210.3212} \BibitemShut {NoStop}%
\bibitem [{\citenamefont {Liu}\ \emph {et~al.}(2013)\citenamefont {Liu},
  \citenamefont {Zhou}, \citenamefont {Wang}, \citenamefont {Liu},
  \citenamefont {He}, \citenamefont {Li},\ and\ \citenamefont {Xu}}]{Liu2013}%
  \BibitemOpen
  \bibfield  {author} {\bibinfo {author} {\bibfnamefont {J.}~\bibnamefont
  {Liu}}, \bibinfo {author} {\bibfnamefont {Y.}~\bibnamefont {Zhou}}, \bibinfo
  {author} {\bibfnamefont {W.}~\bibnamefont {Wang}}, \bibinfo {author}
  {\bibfnamefont {R.-f.}\ \bibnamefont {Liu}}, \bibinfo {author} {\bibfnamefont
  {K.}~\bibnamefont {He}}, \bibinfo {author} {\bibfnamefont {F.-l.}\
  \bibnamefont {Li}}, \ and\ \bibinfo {author} {\bibfnamefont {Z.}~\bibnamefont
  {Xu}},\ }\href {\doibase 10.1364/OE.21.019209} {\bibfield  {journal}
  {\bibinfo  {journal} {Optics express}\ }\textbf {\bibinfo {volume} {21}},\
  \bibinfo {pages} {864} (\bibinfo {year} {2013})}\BibitemShut {NoStop}%
\bibitem [{\citenamefont {Parigi}\ \emph {et~al.}(2007)\citenamefont {Parigi},
  \citenamefont {Zavatta}, \citenamefont {Kim},\ and\ \citenamefont
  {Bellini}}]{Parigi2007}%
  \BibitemOpen
  \bibfield  {author} {\bibinfo {author} {\bibfnamefont {V.}~\bibnamefont
  {Parigi}}, \bibinfo {author} {\bibfnamefont {A.}~\bibnamefont {Zavatta}},
  \bibinfo {author} {\bibfnamefont {M.}~\bibnamefont {Kim}}, \ and\ \bibinfo
  {author} {\bibfnamefont {M.}~\bibnamefont {Bellini}},\ }\href {\doibase
  10.1126/science.1146204} {\bibfield  {journal} {\bibinfo  {journal} {Science
  (New York, N.Y.)}\ }\textbf {\bibinfo {volume} {317}},\ \bibinfo {pages}
  {1890} (\bibinfo {year} {2007})}\BibitemShut {NoStop}%
\bibitem [{\citenamefont {Wenger}\ \emph {et~al.}(2004)\citenamefont {Wenger},
  \citenamefont {Tualle-Brouri},\ and\ \citenamefont {Grangier}}]{Wenger2004}%
  \BibitemOpen
  \bibfield  {author} {\bibinfo {author} {\bibfnamefont {J.}~\bibnamefont
  {Wenger}}, \bibinfo {author} {\bibfnamefont {R.}~\bibnamefont
  {Tualle-Brouri}}, \ and\ \bibinfo {author} {\bibfnamefont {P.}~\bibnamefont
  {Grangier}},\ }\href {\doibase 10.1103/PhysRevLett.92.153601} {\bibfield
  {journal} {\bibinfo  {journal} {Physical Review Letters}\ }\textbf {\bibinfo
  {volume} {92}},\ \bibinfo {pages} {153601} (\bibinfo {year}
  {2004})}\BibitemShut {NoStop}%
\bibitem [{\citenamefont {Xiang}\ \emph {et~al.}(2010)\citenamefont {Xiang},
  \citenamefont {Ralph}, \citenamefont {Lund}, \citenamefont {Walk},\ and\
  \citenamefont {Pryde}}]{Xiang2010}%
  \BibitemOpen
  \bibfield  {author} {\bibinfo {author} {\bibfnamefont {G.~Y.}\ \bibnamefont
  {Xiang}}, \bibinfo {author} {\bibfnamefont {T.~C.}\ \bibnamefont {Ralph}},
  \bibinfo {author} {\bibfnamefont {A.~P.}\ \bibnamefont {Lund}}, \bibinfo
  {author} {\bibfnamefont {N.}~\bibnamefont {Walk}}, \ and\ \bibinfo {author}
  {\bibfnamefont {G.~J.}\ \bibnamefont {Pryde}},\ }\href {\doibase
  10.1038/nphoton.2010.35} {\bibfield  {journal} {\bibinfo  {journal} {Nature
  Photonics}\ }\textbf {\bibinfo {volume} {4}},\ \bibinfo {pages} {316}
  (\bibinfo {year} {2010})},\ \Eprint {http://arxiv.org/abs/0907.3638}
  {arXiv:0907.3638} \BibitemShut {NoStop}%
\bibitem [{\citenamefont {Zavatta}\ \emph {et~al.}(2008)\citenamefont
  {Zavatta}, \citenamefont {Parigi}, \citenamefont {Kim},\ and\ \citenamefont
  {Bellini}}]{Zavatta2008}%
  \BibitemOpen
  \bibfield  {author} {\bibinfo {author} {\bibfnamefont {A.}~\bibnamefont
  {Zavatta}}, \bibinfo {author} {\bibfnamefont {V.}~\bibnamefont {Parigi}},
  \bibinfo {author} {\bibfnamefont {M.~S.}\ \bibnamefont {Kim}}, \ and\
  \bibinfo {author} {\bibfnamefont {M.}~\bibnamefont {Bellini}},\ }\href
  {\doibase 10.1088/1367-2630/10/12/123006} {\bibfield  {journal} {\bibinfo
  {journal} {New Journal of Physics}\ }\textbf {\bibinfo {volume} {10}},\
  \bibinfo {pages} {123006} (\bibinfo {year} {2008})}\BibitemShut {NoStop}%
\bibitem [{\citenamefont {Allevi}\ \emph {et~al.}(2010)\citenamefont {Allevi},
  \citenamefont {Andreoni}, \citenamefont {Bondani}, \citenamefont {Genoni},\
  and\ \citenamefont {Olivares}}]{Allevi2010}%
  \BibitemOpen
  \bibfield  {author} {\bibinfo {author} {\bibfnamefont {A.}~\bibnamefont
  {Allevi}}, \bibinfo {author} {\bibfnamefont {A.}~\bibnamefont {Andreoni}},
  \bibinfo {author} {\bibfnamefont {M.}~\bibnamefont {Bondani}}, \bibinfo
  {author} {\bibfnamefont {M.~G.}\ \bibnamefont {Genoni}}, \ and\ \bibinfo
  {author} {\bibfnamefont {S.}~\bibnamefont {Olivares}},\ }\href {\doibase
  10.1103/PhysRevA.82.013816} {\bibfield  {journal} {\bibinfo  {journal}
  {Physical Review A - Atomic, Molecular, and Optical Physics}\ }\textbf
  {\bibinfo {volume} {82}},\ \bibinfo {pages} {013816} (\bibinfo {year}
  {2010})},\ \Eprint {http://arxiv.org/abs/1003.5840} {arXiv:1003.5840}
  \BibitemShut {NoStop}%
\bibitem [{\citenamefont {Zhai}\ \emph {et~al.}(2013)\citenamefont {Zhai},
  \citenamefont {Becerra}, \citenamefont {Glebov}, \citenamefont {Wen},
  \citenamefont {Lita}, \citenamefont {Calkins}, \citenamefont {Gerrits},
  \citenamefont {Fan}, \citenamefont {Nam},\ and\ \citenamefont
  {Migdall}}]{Zhai2013}%
  \BibitemOpen
  \bibfield  {author} {\bibinfo {author} {\bibfnamefont {Y.}~\bibnamefont
  {Zhai}}, \bibinfo {author} {\bibfnamefont {F.~E.}\ \bibnamefont {Becerra}},
  \bibinfo {author} {\bibfnamefont {B.~L.}\ \bibnamefont {Glebov}}, \bibinfo
  {author} {\bibfnamefont {J.}~\bibnamefont {Wen}}, \bibinfo {author}
  {\bibfnamefont {A.~E.}\ \bibnamefont {Lita}}, \bibinfo {author}
  {\bibfnamefont {B.}~\bibnamefont {Calkins}}, \bibinfo {author} {\bibfnamefont
  {T.}~\bibnamefont {Gerrits}}, \bibinfo {author} {\bibfnamefont
  {J.}~\bibnamefont {Fan}}, \bibinfo {author} {\bibfnamefont {S.~W.}\
  \bibnamefont {Nam}}, \ and\ \bibinfo {author} {\bibfnamefont
  {A.}~\bibnamefont {Migdall}},\ }\href {\doibase 10.1364/OL.38.002171}
  {\bibfield  {journal} {\bibinfo  {journal} {Optics letters}\ }\textbf
  {\bibinfo {volume} {38}},\ \bibinfo {pages} {2171} (\bibinfo {year}
  {2013})}\BibitemShut {NoStop}%
\bibitem [{\citenamefont {{Hashemi Rafsanjani}}\ \emph
  {et~al.}(2017)\citenamefont {{Hashemi Rafsanjani}}, \citenamefont
  {Mirhosseini}, \citenamefont {Maga{\~{n}}a-Loaiza}, \citenamefont {Gard},
  \citenamefont {Birrittella}, \citenamefont {Koltenbah}, \citenamefont
  {Parazzoli}, \citenamefont {Capron}, \citenamefont {Gerry}, \citenamefont
  {Dowling},\ and\ \citenamefont {Boyd}}]{Rafsanjani2017}%
  \BibitemOpen
  \bibfield  {author} {\bibinfo {author} {\bibfnamefont {S.~M.}\ \bibnamefont
  {{Hashemi Rafsanjani}}}, \bibinfo {author} {\bibfnamefont {M.}~\bibnamefont
  {Mirhosseini}}, \bibinfo {author} {\bibfnamefont {O.~S.}\ \bibnamefont
  {Maga{\~{n}}a-Loaiza}}, \bibinfo {author} {\bibfnamefont {B.~T.}\
  \bibnamefont {Gard}}, \bibinfo {author} {\bibfnamefont {R.}~\bibnamefont
  {Birrittella}}, \bibinfo {author} {\bibfnamefont {B.~E.}\ \bibnamefont
  {Koltenbah}}, \bibinfo {author} {\bibfnamefont {C.~G.}\ \bibnamefont
  {Parazzoli}}, \bibinfo {author} {\bibfnamefont {B.~A.}\ \bibnamefont
  {Capron}}, \bibinfo {author} {\bibfnamefont {C.~C.}\ \bibnamefont {Gerry}},
  \bibinfo {author} {\bibfnamefont {J.~P.}\ \bibnamefont {Dowling}}, \ and\
  \bibinfo {author} {\bibfnamefont {R.~W.}\ \bibnamefont {Boyd}},\ }\href
  {\doibase 10.1364/OPTICA.4.000487} {\bibfield  {journal} {\bibinfo  {journal}
  {Optica}\ }\textbf {\bibinfo {volume} {4}},\ \bibinfo {pages} {487} (\bibinfo
  {year} {2017})},\ \Eprint {http://arxiv.org/abs/1605.05424}
  {arXiv:1605.05424} \BibitemShut {NoStop}%
\bibitem [{\citenamefont {Bogdanov}\ \emph
  {et~al.}(2016{\natexlab{a}})\citenamefont {Bogdanov}, \citenamefont
  {Bogdanova}, \citenamefont {Katamadze}, \citenamefont {Avosopyants},\ and\
  \citenamefont {Lukichev}}]{Bogdanov2016b}%
  \BibitemOpen
  \bibfield  {author} {\bibinfo {author} {\bibfnamefont {Yu.~I.}\ \bibnamefont
  {Bogdanov}}, \bibinfo {author} {\bibfnamefont {N.~A.}\ \bibnamefont
  {Bogdanova}}, \bibinfo {author} {\bibfnamefont {K.~G.}\ \bibnamefont
  {Katamadze}}, \bibinfo {author} {\bibfnamefont {G.~V.}\ \bibnamefont
  {Avosopyants}}, \ and\ \bibinfo {author} {\bibfnamefont {V.~F.}\ \bibnamefont
  {Lukichev}},\ }\href {\doibase 10.3103/S8756699016050095} {\bibfield
  {journal} {\bibinfo  {journal} {Optoelectronics, Instrumentation and Data
  Processing}\ }\textbf {\bibinfo {volume} {52}},\ \bibinfo {pages} {475}
  (\bibinfo {year} {2016}{\natexlab{a}})}\BibitemShut {NoStop}%
\bibitem [{\citenamefont {Scully}\ and\ \citenamefont
  {Zubairy}(2001)}]{Scully2001}%
  \BibitemOpen
  \bibfield  {author} {\bibinfo {author} {\bibfnamefont {M.~O.}\ \bibnamefont
  {Scully}}\ and\ \bibinfo {author} {\bibfnamefont {M.~S.}\ \bibnamefont
  {Zubairy}},\ }\href@noop {} {\emph {\bibinfo {title} {{Quantum Optics}}}}\
  (\bibinfo  {publisher} {Cambridge University Press},\ \bibinfo {address}
  {Cambridge},\ \bibinfo {year} {2001})\ p.\ \bibinfo {pages} {630}\BibitemShut
  {NoStop}%
\bibitem [{\citenamefont {Mandel}\ and\ \citenamefont
  {Wolf}(1995)}]{Mandel1995}%
  \BibitemOpen
  \bibfield  {author} {\bibinfo {author} {\bibfnamefont {L.}~\bibnamefont
  {Mandel}}\ and\ \bibinfo {author} {\bibfnamefont {E.}~\bibnamefont {Wolf}},\
  }\href
  {http://books.google.com/books?hl=en{\&}lr={\&}id=FeBix14iM70C{\&}oi=fnd{\&}pg=PR25{\&}dq=Optical+Coherence+and+Quantum+Optics{\&}ots=tgog6{\_}-AxT{\&}sig=7{\_}1nwkK0DOONN5dE0bXy4DKlkBI
  http://books.google.com/books?hl=en{\&}lr={\&}id=FeBix14iM70C{\&}o} {\emph
  {\bibinfo {title} {{Optical coherence and quantum optics}}}},\ \bibinfo
  {edition} {1st}\ ed.\ (\bibinfo  {publisher} {Cambridge University Press},\
  \bibinfo {address} {Cambridge},\ \bibinfo {year} {1995})\ p.\ \bibinfo
  {pages} {1167}\BibitemShut {NoStop}%
\bibitem [{\citenamefont {Ghiu}\ \emph {et~al.}(2014)\citenamefont {Ghiu},
  \citenamefont {Marian},\ and\ \citenamefont {Marian}}]{Ghiu2014}%
  \BibitemOpen
  \bibfield  {author} {\bibinfo {author} {\bibfnamefont {I.}~\bibnamefont
  {Ghiu}}, \bibinfo {author} {\bibfnamefont {P.}~\bibnamefont {Marian}}, \ and\
  \bibinfo {author} {\bibfnamefont {T.~A.}\ \bibnamefont {Marian}},\ }\href
  {\doibase 10.1088/0031-8949/2014/T160/014014} {\bibfield  {journal} {\bibinfo
   {journal} {Physica Scripta}\ }\textbf {\bibinfo {volume} {T160}},\ \bibinfo
  {pages} {014014} (\bibinfo {year} {2014})}\BibitemShut {NoStop}%
\bibitem [{\citenamefont {Martienssen}(1964)}]{Martienssen1964}%
  \BibitemOpen
  \bibfield  {author} {\bibinfo {author} {\bibfnamefont {W.}~\bibnamefont
  {Martienssen}},\ }\href {\doibase 10.1119/1.1970023} {\bibfield  {journal}
  {\bibinfo  {journal} {American Journal of Physics}\ }\textbf {\bibinfo
  {volume} {32}},\ \bibinfo {pages} {919} (\bibinfo {year} {1964})}\BibitemShut
  {NoStop}%
\bibitem [{\citenamefont {Arecchi}(1965)}]{Arecchi1965}%
  \BibitemOpen
  \bibfield  {author} {\bibinfo {author} {\bibfnamefont {F.~T.}\ \bibnamefont
  {Arecchi}},\ }\href {\doibase 10.1103/PhysRevLett.15.912} {\bibfield
  {journal} {\bibinfo  {journal} {Physical Review Letters}\ }\textbf {\bibinfo
  {volume} {15}},\ \bibinfo {pages} {912} (\bibinfo {year} {1965})}\BibitemShut
  {NoStop}%
\bibitem [{\citenamefont {Leonhardt}\ and\ \citenamefont
  {Paul}(1995)}]{Leonhardt1995}%
  \BibitemOpen
  \bibfield  {author} {\bibinfo {author} {\bibfnamefont {U.}~\bibnamefont
  {Leonhardt}}\ and\ \bibinfo {author} {\bibfnamefont {H.}~\bibnamefont
  {Paul}},\ }\href {\doibase 10.1016/0079-6727(94)00007-L} {\bibfield
  {journal} {\bibinfo  {journal} {Progress in Quantum Electronics}\ }\textbf
  {\bibinfo {volume} {19}},\ \bibinfo {pages} {89} (\bibinfo {year}
  {1995})}\BibitemShut {NoStop}%
\bibitem [{\citenamefont {Kumar}\ \emph {et~al.}(2012)\citenamefont {Kumar},
  \citenamefont {Barrios}, \citenamefont {MacRae}, \citenamefont {Cairns},
  \citenamefont {Huntington},\ and\ \citenamefont {Lvovsky}}]{Kumar2012}%
  \BibitemOpen
  \bibfield  {author} {\bibinfo {author} {\bibfnamefont {R.}~\bibnamefont
  {Kumar}}, \bibinfo {author} {\bibfnamefont {E.}~\bibnamefont {Barrios}},
  \bibinfo {author} {\bibfnamefont {A.}~\bibnamefont {MacRae}}, \bibinfo
  {author} {\bibfnamefont {E.}~\bibnamefont {Cairns}}, \bibinfo {author}
  {\bibfnamefont {E.}~\bibnamefont {Huntington}}, \ and\ \bibinfo {author}
  {\bibfnamefont {A.}~\bibnamefont {Lvovsky}},\ }\href {\doibase
  10.1016/j.optcom.2012.07.103} {\bibfield  {journal} {\bibinfo  {journal}
  {Optics Communications}\ }\textbf {\bibinfo {volume} {285}},\ \bibinfo
  {pages} {5259} (\bibinfo {year} {2012})}\BibitemShut {NoStop}%
\bibitem [{\citenamefont {Yamamoto}\ \emph {et~al.}(2007)\citenamefont
  {Yamamoto}, \citenamefont {Yamamura}, \citenamefont {Sato}, \citenamefont
  {Kamakura}, \citenamefont {Ota}, \citenamefont {Suzuki},\ and\ \citenamefont
  {Ohsuka}}]{Yamamoto2007}%
  \BibitemOpen
  \bibfield  {author} {\bibinfo {author} {\bibfnamefont {K.}~\bibnamefont
  {Yamamoto}}, \bibinfo {author} {\bibfnamefont {K.}~\bibnamefont {Yamamura}},
  \bibinfo {author} {\bibfnamefont {K.}~\bibnamefont {Sato}}, \bibinfo {author}
  {\bibfnamefont {S.}~\bibnamefont {Kamakura}}, \bibinfo {author}
  {\bibfnamefont {T.}~\bibnamefont {Ota}}, \bibinfo {author} {\bibfnamefont
  {H.}~\bibnamefont {Suzuki}}, \ and\ \bibinfo {author} {\bibfnamefont
  {S.}~\bibnamefont {Ohsuka}},\ }in\ \href {\doibase
  10.1109/NSSMIC.2007.4437286} {\emph {\bibinfo {booktitle} {2007 IEEE Nuclear
  Science Symposium Conference Record}}}\ (\bibinfo  {publisher} {IEEE},\
  \bibinfo {year} {2007})\ pp.\ \bibinfo {pages} {1511--1515}\BibitemShut
  {NoStop}%
\bibitem [{\citenamefont {Neergaard-Nielsen}\ \emph {et~al.}(2006)\citenamefont
  {Neergaard-Nielsen}, \citenamefont {Nielsen}, \citenamefont {Hettich},
  \citenamefont {M{\o}lmer},\ and\ \citenamefont
  {Polzik}}]{Neergaard-Nielsen2006}%
  \BibitemOpen
  \bibfield  {author} {\bibinfo {author} {\bibfnamefont {J.~S.}\ \bibnamefont
  {Neergaard-Nielsen}}, \bibinfo {author} {\bibfnamefont {B.~M.}\ \bibnamefont
  {Nielsen}}, \bibinfo {author} {\bibfnamefont {C.}~\bibnamefont {Hettich}},
  \bibinfo {author} {\bibfnamefont {K.}~\bibnamefont {M{\o}lmer}}, \ and\
  \bibinfo {author} {\bibfnamefont {E.~S.}\ \bibnamefont {Polzik}},\ }\href
  {\doibase 10.1103/PhysRevLett.97.083604} {\bibfield  {journal} {\bibinfo
  {journal} {Physical Review Letters}\ }\textbf {\bibinfo {volume} {97}},\
  \bibinfo {pages} {1} (\bibinfo {year} {2006})},\ \Eprint
  {http://arxiv.org/abs/0602198} {arXiv:0602198 [quant-ph]} \BibitemShut
  {NoStop}%
\bibitem [{\citenamefont {Riel{\"{a}}nder}\ \emph {et~al.}(2014)\citenamefont
  {Riel{\"{a}}nder}, \citenamefont {Kutluer}, \citenamefont {Ledingham},
  \citenamefont {G{\"{u}}ndoǧan}, \citenamefont {Fekete}, \citenamefont
  {Mazzera},\ and\ \citenamefont {{De Riedmatten}}}]{Rielander2014}%
  \BibitemOpen
  \bibfield  {author} {\bibinfo {author} {\bibfnamefont {D.}~\bibnamefont
  {Riel{\"{a}}nder}}, \bibinfo {author} {\bibfnamefont {K.}~\bibnamefont
  {Kutluer}}, \bibinfo {author} {\bibfnamefont {P.~M.}\ \bibnamefont
  {Ledingham}}, \bibinfo {author} {\bibfnamefont {M.}~\bibnamefont
  {G{\"{u}}ndoǧan}}, \bibinfo {author} {\bibfnamefont {J.}~\bibnamefont
  {Fekete}}, \bibinfo {author} {\bibfnamefont {M.}~\bibnamefont {Mazzera}}, \
  and\ \bibinfo {author} {\bibfnamefont {H.}~\bibnamefont {{De Riedmatten}}},\
  }\href {\doibase 10.1103/PhysRevLett.112.040504} {\bibfield  {journal}
  {\bibinfo  {journal} {Physical Review Letters}\ }\textbf {\bibinfo {volume}
  {112}},\ \bibinfo {pages} {1} (\bibinfo {year} {2014})},\ \Eprint
  {http://arxiv.org/abs/1310.8261} {arXiv:1310.8261} \BibitemShut {NoStop}%
\bibitem [{\citenamefont {Lvovsky}(2004)}]{Lvovsky2004}%
  \BibitemOpen
  \bibfield  {author} {\bibinfo {author} {\bibfnamefont {A.~I.}\ \bibnamefont
  {Lvovsky}},\ }\href {\doibase 10.1088/1464-4266/6/6/014} {\bibfield
  {journal} {\bibinfo  {journal} {Journal of Optics B: Quantum and
  Semiclassical Optics}\ }\textbf {\bibinfo {volume} {6}},\ \bibinfo {pages}
  {S556} (\bibinfo {year} {2004})},\ \Eprint {http://arxiv.org/abs/0311097}
  {arXiv:0311097 [quant-ph]} \BibitemShut {NoStop}%
\bibitem [{\citenamefont {Bogdanov}\ \emph
  {et~al.}(2016{\natexlab{b}})\citenamefont {Bogdanov}, \citenamefont
  {Avosopyants}, \citenamefont {Belinskii}, \citenamefont {Katamadze},
  \citenamefont {Kulik},\ and\ \citenamefont {Lukichev}}]{Bogdanov2016}%
  \BibitemOpen
  \bibfield  {author} {\bibinfo {author} {\bibfnamefont {Yu.~I.}\ \bibnamefont
  {Bogdanov}}, \bibinfo {author} {\bibfnamefont {G.~V.}\ \bibnamefont
  {Avosopyants}}, \bibinfo {author} {\bibfnamefont {L.~V.}\ \bibnamefont
  {Belinskii}}, \bibinfo {author} {\bibfnamefont {K.~G.}\ \bibnamefont
  {Katamadze}}, \bibinfo {author} {\bibfnamefont {S.~P.}\ \bibnamefont
  {Kulik}}, \ and\ \bibinfo {author} {\bibfnamefont {V.~F.}\ \bibnamefont
  {Lukichev}},\ }\href {\doibase 10.1134/S1063776116070025} {\bibfield
  {journal} {\bibinfo  {journal} {Journal of Experimental and Theoretical
  Physics}\ }\textbf {\bibinfo {volume} {123}},\ \bibinfo {pages} {212}
  (\bibinfo {year} {2016}{\natexlab{b}})}\BibitemShut {NoStop}%
\end{thebibliography}

%
\end{document}